\documentclass[aps,10pt,pre,twocolumn,amsmath,amssymb,amsfonts]{revtex4-1}

\usepackage{graphicx}

\def\Eq#1{Eq.(\ref{#1})}
\def\Eqs#1{Eqs.(\ref{#1})}
\def\Fig#1{Fig.~\ref{#1}}

\def\no{\nonumber\\}

\def\>{\rangle}
\def\<{\langle}
\def\rr{\rangle\!\rangle}
\def\lla{\langle\!\langle}

\def\adg{a^\dagger}
\def\Adg{A^\dagger}
\def\bdg{b^\dagger}
\def\half{\tfrac{1}{2}}
\def\Udg{U^\dg}
\def\dg{\dagger}
\def\lam{\lambda}
\def\w{\omega}
\def\wq{\omega_q}
\def\wo{\omega_0}
\def\phio{{\phi_0}}

\def\del{\delta}
\def\gam{\gamma}
\def\Del{\Delta}
\def\al{\alpha}
\def\bt{\beta}

\def\nb{\bar{n}}
\def\d{\partial}

\begin{document}

\title{Attenuation of excitation decay rate due to collective effect}

\author{B.~A.~Tay}
\email{BuangAnn.Tay@nottingham.edu.my}
\affiliation{Foundation Studies, Faculty of Engineering, University of Nottingham Malaysia Campus, Jalan Broga, 43500 Semenyih, Selangor, Malaysia}

\date{\today}

\begin{abstract}

We study a series of $N$ oscillators each coupled to its nearest neighbours, and linearly to a phonon field through the oscillator's number operator. We show that the Hamiltonian of a pair of adjacent oscillators, or a dimer, within the series of oscillators can be transformed to a form in which they are collectively coupled to the phonon field as a composite unit. In the weak coupling and rotating-wave approximation, the system behaves effectively like the trilinear boson model in the one excitation subspace of the dimer subsystem. The reduced dynamics of the one excitation subspace of the dimer subsystem coupled weakly to a phonon bath is similar to that of a two-level system, with a metastable state against the vacuum. The decay constant of the subsystem is proportional to the dephasing rate of the individual oscillator in a phonon bath, attenuated by a factor that depends on site asymmetry, intersite coupling and the resonance frequency between the transformed oscillator modes, or excitons. As a result of the collective effect, the excitation relaxation lifetime is prolonged over the dephasing lifetime of an individual oscillator coupled to the same bath.

\end{abstract}


\maketitle

\section{Introduction}

When the individual members of a group of oscillators assumed to be unrelated to each other are coupled separately to a common radiation field that is coherent, the field will drive these oscillators to radiate in a coherent fashion, resulting in an amplification of stimulated radiation. If this group of oscillators is treated as a single unit, interesting collective aspects such as superradiance will emerge \cite{Dicke54}, even if the field is not coherent.

Collective effects can also be introduced explicitly to the system by coupling the oscillators to each other, and in the simplest case, to their nearest neighbours only \cite{Frohlich52,Holstein59}. The pair of nearest neighbours, or a dimer, forms the smallest collective unit within the group. When the coupling between the oscillator is strong enough, the dimer can be regarded as a composite unit collectively coupled to the field, and its simpler dynamics gives us insights into the collective effects of the group of oscillators.

The system we consider is used to describe the transfer of energy in the form of electronic excitations in light-harvesting complex in photosynthetic systems \cite{[{}][{, and references therein.}]May11}, the transfer of vibrational energy of the amide-I bonds in peptide groups along alpha-helix protein chains \cite{Davydov79,Davydov90}, and the formation and transfer of polarons in deformable media \cite{Frohlich52,Holstein59,Brown86}. In contrast to the common practice of introducing the excitonic basis from the outset to the coupled oscillators \cite{May11,IshizakiJCP09a,*IshizakiJCP09b}, we first carry out a transformation on the field basis \cite{Lee53,Yarkony76,Brown86}, and only after that do we introduce the exciton basis. Apart from introducing the reorganization energy, the transformation on the field replaces the original oscillator-field interaction by a new one in which the pair of adjacent oscillators are collectively coupled to the field.

In the weak coupling and rotating-wave approximation \cite{Cohen-Tannoudji}, the transformed system then reduces to the trilinear boson model \cite{Louisell61,Glauber67,Walls70}. The quantum Markovian master equation of this model can be solved analytically \cite{Tay13}. The reduced dynamics of the dimer subsystem permits a set of metastable states, and possesses a longer relaxation lifetime than the dephasing time of the individual oscillator coupled to the field in a similar way.

The results give us interesting aspects behind the mechanism of excitation energy transfer in photosynthetic system \cite{Engel07,IshizakiPNAS09,Ishizaki10}, and may play a role in the formation of solitons and quantum thermal sound modes in molecular chains \cite{Davydov79,Davydov90,Petrosky09}. Closely related results were also found in the spin-boson model applied to photosynthetic system \cite{Pachon11}, double quantum dot charge qubit system \cite{Vorojtsov05}, and by taking the coherent superposition of excitations and vibrational states into consideration \cite{Christensson12}.

\section{The Hamiltonian}

The Hamiltonian of the system is \cite{May11,Davydov79,Davydov90,Frohlich52,Holstein59,Brown86}
\begin{align}   \label{HN}
    H_N&=H_\text{osc}+ \sum_{q} \wq \bdg_q b_q \no
    &\quad + \sum_q\sum_{m=1}^N  \wq \chi_q^{(m)} \adg_m a_m (b_{-q}+\bdg_q)\,,\\
    H_\text{osc}&=\sum_{m=1}^N  \w_m \adg_m a_m+\sum_{m=1}^{N-1} J_{m,m+1}(\adg_m a_{m+1}+\adg_{m+1} a_m)\,,\label{Hosc}
\end{align}
with units $\hbar=c=1$. This Hamiltonian describes the transfer of excitations between the oscillators. The excitation at site $m$ is represented by an oscillator with creation and annihilation operators, $\adg_m$, and $a_m$, respectively, with frequency or site energy $\w_m$. The oscillators are coupled to their adjacent neighbours with strength $J_{m,m+1}$. When the underlying molecules at each site displace away from their equilibrium positions, they give rise to phonon (field) modes,  created and annihilated by operators $\bdg_q$ and $b_q$, respectively, where $q=\w_q/v$ is the wave vector, and $v$ is the speed of sound. The operators satisfy the usual commutation relation, $[a_m,\adg_{m'}]=\del_{m,m'}$, and similarly for $b_q, \bdg_q$, whereas $a_m, \adg_m$ and $b_q, \bdg_q$ mutually commute. The excitations are coupled to the phonons linearly and modulated by the site energy $\wq \chi_q^{(m)}$, with a $q$-dependent dimensionless coupling constant $\chi_q^{(m)}$ at site $m$. To ensure the hermiticity of the Hamiltonian, $\chi_q^{(m)}$ is required to satisfy the condition $\chi_q^{(m)*}=\chi_{-q}^{(m)}$, where $*$ denotes complex conjugation.

There are two modes of excitation energy transfer in this system \cite{May11}, i.e., the incoherent hopping of excitations between sites described by the F\"{o}rster theory and the coherent wavelike energy transfer over multiple sites in terms of excitons described by the master equation. The latter description is more appropriate when the intersite coupling $J_{12}$ is strong. This is the mode of energy transfer that we are interested in when the collective effect between the oscillators becomes more prominent due to strong intersite coupling.

\subsection{Transformation in phonon basis}

We first subject the Hamiltonian to the unitary transformation \cite{Lee53,Yarkony76,Brown86}
\begin{align}   \label{U1}
    U=\exp\bigg[-\sum_q \sum^N_{m=1}  \chi_q^{(m)} \adg_m a_m (b_{-q}-\bdg_q)\bigg]\,.
\end{align}
By labeling the transformed operator as $O'\equiv UO\Udg$, we find that
\begin{align}   \label{U1ab}
    a'_m&=a_m  \exp\bigg[\sum_q  \chi_q^{(m)}(b_{-q}-\bdg_q)\bigg]\,,\\
    b'_q&=b_q-\sum_{m=1}^N  \chi_q^{(m)} \adg_m a_m\,.
\end{align}
The transformation is sometimes said to dress up the oscillator, where the bare oscillator is now surrounded by a cloud of phonon \cite{Brown86,Gonzalo01}. This process leads to a renormalization on the oscillator's bare frequency \eqref{w'}. As a result, we find that
\begin{align}
    a'^{\dg}_m a'_m&=\adg_m a_m\,,
\end{align}
which implies that the site or number basis labeled by $|n_1,n_2, \cdots\>$ is not altered though the site energies are renormalized, where $n_m$ is the occupation quantum number of the oscillator at site $m$. On the other hand, the phonon basis is altered. The number operator of phonon transforms as
\begin{align} \label{bb}
     b'^{\dg}_q b'_q&= \bdg_q b_q -\sum_{m=1}^N   \chi_q^{(m)} \adg_m a_m (b_{-q}+\bdg_q)\no
            &\quad+\sum_{m=1}^N |\chi_q^{(m)}|^2 \adg_m a_m \no
                &\quad +\sum_{m,m'=1,m\neq m'}^N  \chi_q^{(m)}\chi_q^{(m')*}\adg_m \adg_{m'} a_m a_{m'} \,,
\end{align}

The intersite coupling terms between adjacent oscillators transform into
\begin{align}
   & a'^{\dg}_m a'_{m+1}+ a'^{\dg}_{m+1} a'_m= \adg_m a_{m+1} \exp\bigg[\sum_q \Del^{(m)}_q (b_{-q}-\bdg_q)\bigg]\no
                &\qquad\qquad + \adg_{m+1} a_m \exp\bigg[-\sum_q \Del^{(m)}_q(b_{-q}-\bdg_q)\bigg]\,, \label{aaD}
\end{align}
where
\begin{align}   \label{Dx}
    \Del^{(m)}_q&\equiv\chi_q^{(m+1)}-\chi_q^{(m)} \no
            &\equiv \eta_m \chi_q^{(m)}\,,
\end{align}
denotes the site asymmetry between two adjacent sites. We assume that the difference can be represented by a fraction of $\chi^{(m)}_q$ from a reference site. A small $\eta_m$ can then be used as a dimensionless expansion parameter. In general, $\eta_m$ can be complex.

Under the transformation, the oscillator-phonon interaction becomes
\begin{align}   \label{Hint'}
   a'^\dg_m a'_m (b'_{-q}+b'^{\dg}_q)&= \adg_m a_m(b_{-q}+\bdg_q)-2\chi_q^{m*}\adg_m a_m  \no
   &\quad -2 \sum_{m'=1}^N \chi_q^{(m')*} \adg_m\adg_{m'}  a_m a_{m'} \,.
\end{align}

We will assume that $\Del^{(m)}_q$ is small enough so that the coupling between adjacent oscillators can be approximated by the leading terms in the expansion of the exponentials in \Eq{aaD}. As a result, we obtain the Hamiltonian
\begin{align}   \label{H'}
    &H'_N= H'_\text{osc}+ \sum_{q}  \wq \bdg_q b_q\no
    &\quad -\sum_q \sum_{m,m'=1}^N  \wq \chi_q^{(m)} \chi_q^{m'*}\adg_m \adg_{m'} a_m a_{m'} \no
    &\quad + \sum_q \sum_{m=1}^{N} \eta_m V^{(m)}_q (\adg_m a_{m+1}-\adg_{m+1} a_m)(b_{-q}-\bdg_q)\,,
\end{align}
\begin{align}   \label{Hosc'}
    H'_\text{osc}=\sum_{m=1}^N  \w'_m \adg_m a_m+\sum_{m=1}^{N-1} J_{m,m+1}(\adg_m a_{m+1}+\adg_{m+1} a_m)\,,
\end{align}
where
\begin{align}   \label{V}
    V^{(m)}_q\equiv J_{m,m+1} \chi_q^{(m)}\,.
\end{align}
and the site frequency becomes
\begin{align}   \label{w'}
    \w'_m\equiv\w_m-\sum_q \wq|\chi_q^{(m)}|^2=\w_m-2\lam_m\,,
\end{align}
in which
\begin{align}   \label{reorg}
        \lam_m\equiv\frac{1}{2}\sum_q  \wq|\chi_q^{(m)}|^2
\end{align}
is the reorganization energy \cite{May11,IshizakiJCP09a,*IshizakiJCP09b}. Notice that the original oscillator-phonon interaction cancels out, and is replaced by a new interaction arises from the intersite coupling between adjacent oscillators, see the third line of \Eq{H'}. The term on the second line of \Eq{H'} with $\adg_m \adg_{m'} a_m a_{m'}$ is a many-body term. It can be dropped when we restrict our consideration to the subspace of no more than one excitation \cite{Brown86}. This subspace consists of the site basis
\begin{align}   \label{012}
    |0\>\equiv|0,0,0,\cdots\>\,, \,\, |1\>\equiv|1,0,0,\cdots\>\,, \,\,|2\>\equiv|0,1,0,\cdots\>\,,
\end{align}
and so on, where the positive integer in $|m\>$ denotes an excitation at site $m$.

\subsection{Diagonalization of dimer subsystem}

We will now focus our attention on the dynamics of two adjacent sites, i.e., a dimer subsystem, with the phonon field. Since the adjacent oscillators are coupled to each other, we will later see that they can be effectively viewed as a composite unit collectively coupled to the phonon field. For simplicity, we label the adjacent oscillators generically by 1 and 2, where 1 denotes the oscillator with a greater frequency among the two. Setting $N=2$ in \Eqs{H'} and \eqref{Hosc'}, and dropping operators with index 3 that belongs to another dimer, we denote the resulting system Hamiltonian by $H'$.

The dimer's Hamiltonian
\begin{align}   \label{H'0}
    H'_0 = \w'_1 \adg_1 a_1+\w'_2 \adg_2 a_2+J_{12} (\adg_1 a_2+\adg_2 a_1)
\end{align}
can be diagonalized by a complex rotation through an angle $\phi$
\begin{align}   \label{U2}
    U_\phi=\exp(-i\phi L_2)
\end{align}
along the operator
\begin{align}
    L_2&= \frac{1}{2i}(\adg_1 a_2 -\adg_2 a_1 ) \,.\label{L2}
\end{align}
This operator forms one of the algebra elements of the bosonic representation of the SU(2) \cite{Schwinger}, see App.~\ref{AppSU2}. This rotation does not affect the phonon operators, nor does it affect the oscillator-phonon interaction term that is proportional to $L_2$ as can be seen directly from \Eq{H'}. Subjecting $H'_0$ to the transformation $U_\phi H'_0 \Udg_\phi$, we find that it can be diagonalized by choosing the angle as
\begin{align}   \label{phi}
    \phi_0=\tan^{-1}\frac{-2J_{12}}{ \w'_1-\w'_2}\,,
\end{align}
see App.~\ref{AppSU2}.

Next, we introduce the exciton operators defined by $A_i\equiv \Udg_\phio a_i U_\phio$, and similarly for their Hermitian conjugate, where the index $i=1, 2$ from now on. Note the order of $\Udg_\phio, U_\phio$ in the definition of $A_i$ is different from the previous transformation. We find that
\begin{align} \label{A1}
    A_1&=a_1 \,\cos(\phio/2) - a_2\,\sin(\phio/2) \,,\\
    A_2&=a_1 \sin(\phio/2)+a_2 \cos(\phio/2)\,. \label{A2}
\end{align}
The exciton operators obey the commutation relation $[A_i,\Adg_j]=\del_{ij}$. The exciton basis consists of $|e_i\>\equiv U^\dg_\phio|i\>$, which satisfies the normalization condition $\<e_i|e_j\>=\del_{ij}$. $\Adg_i$ and $A_i$ raises and lowers the exciton states, respectively,
\begin{align}   \label{A|i>}
    \Adg_i|e_0\>=|e_i\>\,, \qquad A_i|e_j\>=\del_{ij}|e_0\>\,.
\end{align}
Notice from \Eq{phi} that in the limit $J_{12}\gg \w'_1-\w'_2$, $\phi_0 \rightarrow \pi/2$, both $A_1, A_2$ consist of equal weights of $a_1, a_2$, see \Eqs{A1} and \eqref{A2}.

In terms of the exciton operators, the system Hamiltonian becomes,
\begin{align}   \label{Hb}
    H'&=H'_0+\sum_{q}  \wq \bdg_q b_q + \eta \sum_q V_q (L_+ b_{-q}+L_-\bdg_q)\no
                        &\quad- \eta \sum_q V_q (L_- b_{-q}+L_+\bdg_q)\,,\\
    H'_0&= \w_+ \Adg_1 A_1+ \w_- \Adg_2 A_2\,, \label{Hb0}
\end{align}
where
\begin{align}   \label{L+}
    L_+&\equiv\Adg_1 A_2 \,, &\quad      L_-&\equiv \Adg_2 A_1\,,
\end{align}
are the raising and lowering operators for the composite system of two excitons, in which both excitons are collectively viewed as a unit. We have omitted the index 1 on $\eta_1$ and $V^{(1)}_q$ for simplicity. Notice that when the site asymmetry vanishes $\eta_m=0$, though $H'$ for a dimer system \eqref{Hb} can be diagonalized up to a many-body term, but not $H'_N$ \eqref{H'} involving all the oscillators, due to the coupling between the constituents of different dimers, in $H'_\text{osc}$ \eqref{Hosc'}.

When we assume $\w'_1>\w'_2$, and choose the angle to lie in the range $-\pi/2\leq \phi_0 \leq \pi/2$, the frequency of exciton 1 remains greater than that of exciton 2, $\w_+>\w_-$. Their frequencies are explicitly given by
\begin{align}   \label{wb1}
    \w_+&\equiv\w'_1\cos^2(\phio/2)+\w'_2\sin^2(\phio/2)- J_{12}\sin\phio\no
    \w_-&\equiv\w'_1\sin^2(\phio/2)+\w'_2\cos^2(\phio/2)+ J_{12}\sin\phio\no
    \w_\pm&=\frac{1}{2}(\w'_1+\w'_2)\pm \frac{1}{2}\sqrt{(\w'_1-\w'_2)^2+4J^2_{12}}\,.
\end{align}
The site basis is related to the exciton basis explicitly by
\begin{align}   \label{e1}
    |1\>&=\cos(\phio/2)|e_1\>+\sin(\phio/2)|e_2\>\,,\\
    |2\>&=-\sin(\phio/2)|e_1\>+\cos(\phio/2)|e_2\>\,,\label{e2}
\end{align}
while the vacuum state remains invariant $|e_0\>=U^\dg_\phio|0\>=|0\>$.

Since $\w_+>\w_-$, the exciton-phonon coupling in the second line of \Eq{Hb} describes virtual processes, i.e., where exciton 1 and phonon are simultaneously excited and created, or simultaneously relaxed and annihilated, respectively. These are fast oscillating terms that average to zero and are usually neglected under the rotating-wave approximation \cite{Cohen-Tannoudji}. If we have the opposite situation $\w_+<\w_-$, the exciton-phonon coupling terms in the first line of \Eq{Hb} will now describe virtual processes and can be dropped instead under the rotating-wave approximation.

As a result of two unitary transformations, we arrive at an effective Hamiltonian where both excitons are collectively coupled to the phonon. It has the same form as the trilinear boson model that is used to describe the processes of parametric amplification and frequency conversion in quantum optics \cite{Louisell61,Glauber67,Walls70}. By restricting our consideration to its one-particle subspace, it is formally the same as the Friedrichs-Lee model \cite{Friedrichs48,Lee} used to study resonances in unstable systems \cite{Friedrichs48,Kallen55,Glaser56,Petrosky91}, and renormalizable field theory \cite{Lee}.

\section{Solution of the Markovian master equation}
\label{SecMME}

The reduced dynamics of a pair of excitons in a thermal bath of phonons interacting through the trilinear boson model has the Kossakowski-Lindblad's form in the weak coupling limit, and can be solved analytically \cite{Tay13}. The exciton subsystem density matrix, $\rho$, evolves according to the equation $\d\rho/\d t=-K\rho$, where \cite{Tay13}
\begin{align}   \label{K}
    K=K_0+K_d
\end{align}
can be decomposed into a unitary part,
\begin{align}   \label{K0}
K_0&\rho
    = i[H'_0,\rho]\,,
\end{align}
and a dissipative part,
\begin{align}  \label{Kd}
    K_d\rho&=-\half  \gam  \nb_0 (2 L_+ \rho L_- - L_-L_+ \rho -\rho L_-L_+ ) \no
            & -\half \gam (\nb_0+1) (2 L_- \rho L_+ - L_+L_- \rho - \rho L_+L_-)   \,.
\end{align}
In the unitary part $K_0$, there are renormalizaions to the exciton frequencies $\w_\pm$, see \Eqs{wtilde}-\eqref{nbk}, that we will ignore in our discussion since they do not affect our results. We have also dropped a many-body term in $K_0$ \cite{Tay13} since it does not contribute to the reduced dynamics in the one excitation subspace we consider. We further assume that the phonon bath has the Bose-Einstein distribution
\begin{align}
     \nb_0\equiv \frac{1}{\exp(\wo/kT)-1} \label{nb0}\,,
\end{align}
where $T$ is the temperature of the phonon bath and $\w_0$ is the resonant frequency between the pair of excitons
\begin{align}   \label{wb0}
    \w_0&\equiv \w_+ -\w_- \no
        &=\sqrt{[\w_1-\w_2+2\lam_1 |\eta|(2 \cos\theta +|\eta|)]^2+4J^2_{12}}\,,
\end{align}
in which the angle $\theta$ is defined by
\begin{align}   \label{theta}
    \cos \theta \equiv \frac{\text{Re}(\eta)}{|\eta|}\,.
\end{align}
To obtain \Eq{wb0}, we have made use of the relation
\begin{align}   \label{lam12}
    \lam_2-\lam_1=\lam_1|\eta|(2 \cos\theta+|\eta|)\,,
\end{align}
deduced from \Eq{reorg} and $\chi^2_q=(1+\eta)\chi_q^1$ \eqref{Dx}. The decay constant of the reduced dynamics has the form \cite{Tay13}
\begin{align}   \label{gam}
    \gam&\equiv 2\pi |\eta|^2\sum_q |V_q|^2 \del(\w_q-\w_0)\\
    &=\al \gam_d\,,\label{gam=gamd}
\end{align}
where
\begin{align}   \label{al}
        \al&\equiv \left(|\eta| \frac{J_{12}}{\w_0}\right)^2 \,,
\end{align}
and
\begin{align}   \label{gamm}
        \gam_d&\equiv 2\pi \sum_q \w_q^2|\chi_q^{(1)}|^2\del(\w_q-\w_0)
\end{align}
is the dephasing constant of an individual oscillator at site $1$ immersed in a phonon bath with the oscillator-phonon interaction in \Eq{HN}. In the case of zero site asymmetry or vanishing intersite coupling, the exciton and phonon decouple from each other and there is no transition between the exciton states. $\al$ is reflection symmetric with respect to $\theta$ since $\wo$ depends on $\cos\theta$ \eqref{wb0}.

The attenuation factor $\al$ is proportional to the square of the ratio between the time scale of oscillation between the excitons, $1/\w_0$, and the time scale of exciton transfer due to intersite coupling, $1/J_{12}$. As a result of the attenuation, the time scale of exciton decay, $1/\gam$, is prolonged by a factor of $1/\al$ over the time scale of dephasing at the individual site, $1/\gam_d$. In principle, each site has its own value of $\gam_d$ and $\al$. Hence, $\gam$ has different values for different pairs of excitons.

We note that it is common in the studies on this system to introduce the excitonic basis from the beginning prior to the unitary transformations \cite{May11,Pachon11,Plenio12}. Consequently, the Hamiltonian contains additional longitudinal or diagonal term of the form $\sum_{q,m}\w_q\chi_q^{(m)} \Adg_m A_m (b_{-q}+\bdg_q)$ in \Eq{H'}. This term leads to fluctuation in the excitons' energies and gives rise to an additional pure dephasing contribution to the relaxation of the total relaxation rate in the Markovian master equation \cite{Pachon11,Plenio12}. But this contribution does not appear explicitly in our formulation \eqref{gam=gamd}.

In the weak coupling approximation to derive the Markovian master equation, we require the factor $\al$ \eqref{al} to be small. When the intersite coupling is comparable to the excitonic transition energy $\w_0$, such as in the photosynthetic systems discussed in Sec.~\ref{SecPhotosyn}, this condition can still be fulfilled if the site asymmetry $|\eta|$ is relatively small.

In the reduced dynamics of $K$ \eqref{K}, the total excitation quantum number is a constant of motion. The underlying Liouville space therefore separates into disconnected subspaces according to this quantum number \cite{Tay13}. Each subspace evolves independently of each other and behaves like a finite-level system. In particular, the one exciton subspace exhibits a dynamics similar to that of a two-level system. Each subspace separately possesses a nondegerate equilibrium state. They are metastable states before processes that we have ignored so far, such as the virtual processes, return the excitations to the vacuum state. Apart from a prolonged relaxation time arises from the collective effect of the coupled oscillators, the excitation energy transfer in this system is also facilitated by the existence of metastable state in each subspace that further prolongs the excitation lifetime.

When the excitations are fermionic in nature, we can replace the bosonic operators $\Adg_i, A_i$, by the Pauli spin matrices $\sigma^{(i)}_\pm=(\sigma^{(i)}_1\pm i \sigma^{(i)}_2)/2$ for exciton $i$, respectively. They obey the commutation relation $[\sigma^{(i)}_+,\sigma^{(i)}_-]= \sigma^{(i)}_3$, the anti-commutation relation $\{\sigma^{(i)}_+,\sigma^{(j)}_-\}= \del_{ij}I$, and they anti-commute with one another $\{\sigma^{(i)}_\pm,\sigma^{(j)}_\pm\}= 0$. The expressions for the $A_i$s are then valid for the $\sigma^{(i)}$s too.

\section{Applications of the reduced dynamics}

\subsection{Photosynthetic systems}
\label{SecPhotosyn}

In photosynthetic systems, such as the light-harvesting complex in purple bacteria \cite{[{}][{, and references therein.}]Sundstrom99} and the Fenna-Matthews-Olson (FMO) pigment protein complex in green sulfur bacteria \cite{[{}][{, and references therein.}]Adolphs06}, the basic light-harvesting unit consists of a group of bacteriachlorophyll pigments held by an underlying protein structure. Photons from sunlight excite electrons in these pigments. The excitation energy then transfers from one pigment to the others until it finally reaches the reaction center, where charge separation occurs and the excitation energy is kept in chemical compounds. It was found that the energy transfer process in photosynthetic systems is extremely efficient. With the discovery of the long-lived quantum coherence in this system, it was suggested that oscillations in the site populations as a result of wavelike energy transfer increases the probability of energy transfer to the reaction center \cite{Engel07}.

Electronic excitations in the pigments are modeled by the excitations of oscillators at each site. The excitations transfer across the pigments to the reaction center while the process is attenuated by the interactions with phonon modes arise from the underlying protein structure that carries them. It was found that the correlation of the bath modes between different sites are weak \cite{Olbrich11}. This occurs, for example, when the pigments are far apart compared to the bath correlation length. The bath modes at each site can then be treated as independent \cite{Gilmore06,Silbey02}. By treating the bath modes as independent oscillators and expanding the bath operators in terms of their normal modes, we obtain the Hamiltonian in \Eq{HN}. For a low sunlight intensity, we can restrict our consideration to the one excitation subspace only.

In the usual studies on photosynthetic systems, the influence of the bath on the system is often characterized by the spectral density
$J_m(\w)\equiv \sum_q \w_q^2 (\chi_q^{(m)})^2 \del(\w-\w_q)$  \cite{May11},
which is usually assumed to be the same for different sites \cite{IshizakiJCP09a,*IshizakiJCP09b,Plenio12,Adolphs06}. Whereas in our study, we consider $\chi_q^{(m)}$ as site-dependent. The difference in the influence of the bath on different sites is encoded in the site asymmetry parameter $\eta_m$. We can then estimate $|\eta_m|$ using experimental results. In the literature, the site dependence is sometimes considered by assuming that the correlation functions between different sites decay exponentially with respect to intersite distance \cite{Adolphs06,Mukamel11}.

As an example, let us consider the chlorophyll pigments labeled by 1 and 2 in the FMO pigment protein complex of the green sulfur bacteria \textit{Chlorobium tepidium} \cite{Adolphs06}. The pigment 2 with greater site energy is labeled by $i=1$. We note that we are extending our result to photosynthetic systems, even though the intersite coupling, $J_{12}=96 \,\text{cm}^{-1}$, is comparable to the site energy difference, $\w_1-\w_2=120 \,\text{cm}^{-1}$ \cite{Adolphs06}, so far as $|\eta|$ is small enough so that the weak coupling assumption is still valid, see the discussion in Sec.~\ref{SecMME}. We assume the reorganization energy $\lam_1=35 \,\text{cm}^{-1}$ \cite{IshizakiPNAS09,*Ishizaki10}.

\begin{figure}[t]
  \centering
\includegraphics[width=3.2in]{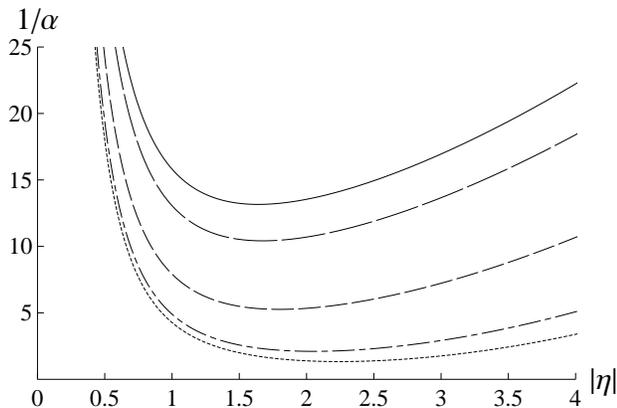}
\caption{$1/\al$ versus $|\eta|$ for a few $\theta$, with parameters $\w_1-\w_2=120 \,\text{cm}^{-1}$, $J_{12}=-96 \,\text{cm}^{-1}$ \cite{Adolphs06}, and $\lam_1=35\,\text{cm}^{-1}$ \cite{IshizakiPNAS09,*Ishizaki10}. Solid, long-dashed, short-dashed, dot-dashed, and dotted lines represent $\theta=0$,  $\pi/4$, $\pi/2$, $3\pi/4$, and $\pi$, respectively.}
\label{fig1}
\end{figure}

In \Fig{fig1}, we plot $1/\al$ against $|\eta|$ for a few values of $\theta$ lying in $0\leq\theta\leq\pi$, and we recalled that $\al$ is reflection symmetric with respect to $\theta$, see \Eqs{al} and \eqref{wb0}. Each curve has a minimum. The second and third row of Table \ref{table1} is a list of their coordinates. The table shows that $1/\al$ are greater than 1 for all curves. Hence, in general the relaxation time scale of this dimer subsystem is longer than the site dephasing time scale $1/\gam_d$ by the factor $1/\al$. We also note that for equal $|\eta|$, greater $\text{Re}(\eta)=|\eta| \cos\theta$ results in longer relaxation lifetime for the excitons.

\begin{table}[t]
\centering
\begin{tabular}{lcccccccccc}
    \hline\hline
    $\theta$ &&0 &&$\pi/4$ &&$\pi/2$ &&$3\pi/4$ &&$\pi$\\
    \hline
        $|\eta|_\text{min}$ &&1.64 &&1.68 &&1.80 &&2.05 &&2.24\\
    $(1/\al)_\text{min}$ &&13.2 &&10.4 &&5.26 &&2.10 &&1.33\\
    $|\eta|$ &&0.71 &&0.63 &&0.53 &&0.47 &&0.45\\
    $\lam_2$ (cm$^{-1}$) &&102 &&80 &&45 &&19 &&11\\
    \hline\hline
\end{tabular}
\caption{Coordinates of the minimum of $1/\al$ for a few $\theta$ in \Fig{fig1}. Using the same parameters in \Fig{fig1}, $\gam_d=50$ fs and $1/\gam_{12}=1100$ fs \cite{Engel10}, the fourth row is the smallest real solution of $|\eta|$ when $1/\al=\gam_d/\gam_{12}=22$, and the last row is $\lam_2$ calculated from \Eq{lam12} in cm$^{-1}$.}
\label{table1}
\end{table}

To get an estimate of $|\eta|$, let us use the experimental values for the relaxation lifetime between pigment 1 and 2, $1/\gam_{12}=1100$ fs \cite{Engel10}, and a site dephasing time of $1/\gam_d=50$ fs \cite{IshizakiPNAS09,*Ishizaki10}, yielding $\gam_d/\gam_{12}=22$. We note that the value $\w_1-\w_2=120$ cm$^{-1}$ we use is slightly different from 160 cm$^{-1}$ in Ref.~\cite{Engel10}. Setting $1/\al=\gam_d/\gam_{12}$ \eqref{gam=gamd} gives a quartic equation in $|\eta|$. The fourth row of Table \ref{table1} lists the smallest real solution of $|\eta|$ for various $\theta$.

Using \Eq{lam12}, we can further estimate $\lam_2$. The results are listed in the last row of Table \ref{table1} for corresponding $|\eta|$ and $\theta$, with $\lam_2$ ranges from 11 to 102 cm$^{-1}$.

In Sec.~\ref{SecMME}, we have mentioned that the excitonic picture for the excitation energy transfer is applicable for strong intersite coupling. In the opposite case when the intersite coupling is much smaller than the difference in the site energies $\w_1-\w_2$, the use of the excitonic picture becomes problematic since energy transfer now moves into the incoherent hopping mode. As an example, let us consider the pigment 1 and 3 of the same protein pigment complex. Pigment 3 with smaller site energy than pigment 1 is labeled by $i=2$. Using $\w_1-\w_2=200\, \text{cm}^{-1}, J_{12}=5.0 \,\text{cm}^{-1}$ \cite{Adolphs06}, we have a small ratio of $J_{12}/(\w_1-\w_2) =0.025$. In the limit where both $\lam_1/(\w_1-\w_2)$ and $J_{12}/(\w_1-\w_2) \rightarrow 0$, we obtain from \Eqs{al} and \eqref{wb0},
\begin{align}   \label{inverselimal}
\frac{1}{\al} \rightarrow \frac{1}{|\eta|^2}\cdot \left(\frac{\w_1-\w_2}{J_{12}}\right)^2\,.
\end{align}
The experimental value for the relaxation lifetime is $1/\gam_{13}=700$ fs \cite{Engel10}. If we assume the same site dephasing constant time scale as in the previous example $1/\gam_d=50$ fs, we obtain $\gam_d/\gam_{13} = 14$. The solution to \Eq{gam=gamd} using \Eq{inverselimal} gives an extremely large $|\eta|=10.7$. This indicates that the excitonic picture is not appropriate in describing excitation energy transfer when $J_{12}/(\w_1-\w_2) \ll 1$.

The time evolution of the site populations can be worked out readily \cite{Tay13}. The results are listed in App.~\ref{AppSol}. We note that a longer relaxation lifetime will naturally lead to a longer oscillations in the site populations. Furthermore, because of the existence of metastable state in the exciton basis, although the correlation component of the density matrix between exciton 1 and 2 vanishes asymptotically \eqref{R12t}, but the real part of the correlation between site 1 and 2 \eqref{r12t} remains finite.

A closely related work \cite{Pachon11} also obtained a longer relaxation lifetime for FMO photosynthetic complexes using the spin-boson model for independent bath oscillators, where the bath is characterized by the spectral density that is Ohmic.

\subsection{Alpha-helix protein molecular chain}

It is interesting to apply the result to the transfer of vibrational energy in alpha-helix protein molecular chain, which consists of a series of amide-I peptide groups. The vibrational energy transfers along the molecular chain through dipole-dipole interactions between the bonds in the groups \cite{Davydov79,Davydov90}. As the vibrational modes (vibrons) travel along the chain, they are modulated by acoustic modes (phonons) created by the displacement of the molecules from their equilibrium positions in the underlying protein chain.

Assuming that the amide-I peptide groups lie on a regular lattice, we assume $\chi_q^{(m)}=\chi_q \exp(imqa)$ \cite{Brown86}, where $a$ is the distance between adjacent peptide groups. In the long wave length limit, $qa\ll 1$, we obtain
\begin{align}   \label{chi-chi}
    \chi_q^{(2)}-\chi_q^{(1)}&= 2i \sin(qa/2) \chi_q e^{3qai/2}\no
                    &\approx iqa\chi_q e^{3qai/2}\,.
\end{align}
Therefore, $|\eta|\approx qa=\w_q a/v$, where $v$ is the speed of sound in the lattice. At resonant frequency $\w_0$, we obtain the decay constant
\begin{align}   \label{gamprotein}
    \gam_\text{hx}&=\al_\text{hx} \gam_{d, \text{hx}}\,,\\
    \al_\text{hx}&\equiv \left(\frac{a}{v} J_{12}\right)^2\,,
\end{align}
where $\gam_{d,\text{hx}}$ is the dephasing rate at the individual site. $\gam_{d,\text{hx}}$ has similar expression as \Eq{gamm}, except that there is no site dependence on $\chi_q$. In this case, $\al_\text{hx}$ is proportional to the square of the ratio between the time scale for sound wave to traverse adjacent sites, $a/v$, and the time scale for the vibrational mode to move from one site to another via intersite coupling, $1/J_{12}$. Using the value $J_{12}=7.8\, \text{cm}^{-1}$, $a=4.5\,${\AA}, and $v=4000\,\text{m/s}$ at physiological temperature $310$ K \cite{Davydov90,Petrosky09}, we obtain a longer relaxation time scale of about $1/\al=36.6$ times that of the dephasing time scale of the individual site in a phonon bath. Such a prolonged relaxation time scale and the existence of metastable in each fixed exciton number subspace may facilitate the formation of solitons \cite{Davydov79,Davydov90} or quantum thermal sound modes in this system \cite{Petrosky09}.

\section{Conclusion}

We show that the Hamiltonian of the dimer subsystem within a series of $N$ oscillators coupled to their nearest neighbours, and to the phonon field produced by the displacement of the underlying molecular structure, can be reduced to the trilinear boson model under the weak coupling and rotating-wave approximation. Due to the collective effect arises from the coupled adjacent oscillators and the phonon bath, there exists metastable states in the reduced dynamics of the dimer subsystem, and a prolonged relaxation lifetime of the excitations over the dephasing time of the individual uncoupled oscillator. These properties can facilitate the transfer of excitation energy in the system, such as in the photosynthetic complexes. They may also play a role in the formation of solitons or quantum thermal sound modes in molecular chains.

\acknowledgments

We acknowledge the insight of an anonymous referee of Ref.~\cite{Tay13} who suggested a possible application of the results to photosynthetic systems that initiated this work. Support by the Ministry of Higher Education, Malaysia (MOHE) under the Fundamental Research Grant Scheme (FRGS), Grant No.~FP009-2011A, is gratefully acknowledged.

\appendix

\section{SU(2) bosonic representation}
\label{AppSU2}

The set of operators
\begin{align}
    L_1 &= \frac{1}{2} (\adg_1 a_2 +\adg_2 a_1 )\,,\label{L1} \\
    L_3&=\frac{1}{2} (\adg_1 a_1-\adg_2 a_2)\,, \label{L3}
\end{align}
and $L_2$ in \Eq{L2} forms the bosonic representation of the algebra of SU(2) \cite{Schwinger}. Under the rotation $U_\phi$ defined in \Eq{U2}, $L_1$ and $L_3$ transform into
\begin{align}
    U_\phi L_1 \Udg_\phi &=L_1\, \cos \phi - L_3\, \sin \phi \,,\\
    U_\phi L_3 \Udg_\phi &=L_1\, \sin \phi + L_3\, \cos \phi \,,
\end{align}
whereas $L_2$ and
\begin{align}   \label{L0}
    L_0 &= \frac{1}{2} (\adg_1 a_1+\adg_2 a_2)\,,
\end{align}
are invariant.

In terms of the $L_i$s, the Hamiltonian of the dimer \eqref{H'0} is
\begin{align}   \label{H'Li}
    H'_0 &= (\w'_1+\w'_2) L_0 +(\w'_1-\w'_2) L_3+2 J_{12} L_1 \,.
\end{align}
The $2J_{12}L_1$ term can be rotated away with the operator \eqref{U2} by choosing the angle $\phio$ \eqref{phi}. We then obtain
\begin{align}   \label{H'Lidiag}
    U_\phio H'_0 \Udg_\phio &= \w_+ \adg_1 a_1+ \w_- \adg_2 a_2 \,.
\end{align}
By substituting the exciton operators \Eqs{A1}-\eqref{A2} into $H'_0$ and $H'$, we obtain \Eqs{Hb}-\eqref{Hb0}.

\section{Renormalized frequencies}
\label{AppRenormFreq}

The renormalized excitonic frequencies due to the influence of the phonon bath are \cite{Tay13}
\begin{align} \label{wtilde}
    \bar{\w}_\pm &\equiv \w_\pm - \del \w_\pm\,,\\
    \del\w_+ &\equiv \sum_k \text{P} \frac{|V_k|^2}{\w_k-\w_0}(\nb_k+1)\,,\\
    \del\w_- &\equiv -\sum_k \text{P} \frac{|V_k|^2}{\w_k-\w_0}\nb_k\,,\\
    \nb_k &\equiv \frac{1}{e^{\w_k\bt}-1}\,.\label{nbk}
\end{align}

\section{Time evolution of populations in the site basis}
\label{AppSol}

To find out the time evolution of the site populations, we need to first solve the Markovian master equation in the exciton basis, and then convert the solution back to the site basis. A density matrix in the one exciton subspace can be written as
\begin{align}   \label{rho012}
    \rho(t)&=\rho_{e_0e_0}(t)|e_0;e_0\rr+\rho_{e_1e_1}(t)|e_1;e_1\rr+\rho_{e_2e_2}(t)|e_2;e_2\rr\no
        & +\big[\rho_{e_0e_1}(t)|e_0;e_1\rr+\rho_{e_0e_2}(t)|e_0;e_2\rr+\rho_{e_1e_2}(t)|e_1;e_2\rr\no
        &\qquad +\text{H.c.}\big]\,,
\end{align}
where we define $|e_i;e_j\rr\equiv|e_i\>\<e_j|$. For completeness, we have included the vacuum state. The solutions to the components of the density matrix are
\begin{align}
        \rho_{e_0e_0}(t)&=\rho_{e_0e_0}(0)\,,\\
       \rho_{e_0e_1}(t)&=\rho_{e_0e_1}(0) e^{-\gam(1+\nb_0)t/2}e^{i\, \w_1 t}\,,\label{gt0}\\
        \rho_{e_0e_2}(t)&=\rho_{e_0e_2}(0) e^{-\gam\nb_0 t/2}e^{i \, \w_2 t}\,,\label{gt0b}\
\end{align}
\begin{align}
        \rho_{e_1e_1}(t)&=\rho_{e_1e_1}(0)e^{-\gam(1+2\nb_0)t}\no
                &\quad +\frac{\nb_0[1-\rho_{e_0e_0}(0)]}{1+2\nb_0} \left(1-e^{-\gam(1+2\nb_0)t}\right)\,,\label{R11t}\\
        \rho_{e_2e_2}(t)&=1-\rho_{e_0e_0}(0)-\rho_{e_1e_1}(t)\,,\\
         \rho_{e_1e_2}(t)&= \rho_{e_1e_2}(0) e^{-(1+2\nb_0)\gam t/2}e^{-i\w_0t}\,.\label{R12t}
\end{align}
The solutions in the site basis defined by the components
\begin{align}   \label{rhoexc}
    \rho_{ij}(t)\equiv\lla i,j|\rho(t)\rr\,,
\end{align}
have the following forms
\begin{align}
        \rho_{00}(t)&= \rho_{e_0e_0}(0)\,,\\
        \rho_{01}(t)&= \rho_{e_0e_1}(t)\,\cos(\phio/2)+\rho_{e_0e_2}(t)\, \sin(\phio/2)\,,
        \end{align}
\begin{align}
        \rho_{02}(t)&= -\rho_{e_0e_1}(t)\,\sin(\phio/2)+\rho_{e_0e_2}(t)\, \cos(\phio/2)\,,\\
        \rho_{11}(t)&= \frac{1}{2}+\left(\rho_{e_1e_1}(t)-\frac{1}{2}\right)\cos\phio
                    +\text{Re}(\rho_{e_1e_2}(t))\sin\phio\,,\label{r11t}
\end{align}
\begin{align}
        \rho_{22}(t)&=1-\rho_{11}(t)\,,\\
         \rho_{12}(t)&=\left(\frac{1}{2}-\rho_{e_1e_1}(t)\right)\sin\phio+\text{Re}(\rho_{e_1e_2}(t))\cos\phio  \no
                    &\quad +i\,\text{Im}(\rho_{e_1e_2}(t)) \,.\label{r12t}
\end{align}
\Eqs{r11t} and \eqref{R12t} indicate that the real part of the inter-exciton correlation component $\text{Re}(\rho_{e_1e_2})$ gives rise to oscillation in the site populations. Since a smaller decay constant of the reduced dynamics $\gam$ leads to a longer relaxation time in the excitonic correlation $\rho_{e_1e_2}$ \eqref{R12t}, the oscillation in the site populations will last longer too.

\providecommand{\noopsort}[1]{}\providecommand{\singleletter}[1]{#1}%

\end{document}